\documentclass{article}
\usepackage{graphicx}
\usepackage{amssymb}
\fontsize{14}{24pt}\selectfont

\title{Scattering problem in deformed space with minimal length}
\author{M.M. Stetsko\footnote{E-mail: mykola@ktf.franko.lviv.ua}\ \ and
V.M. Tkachuk\footnote{E-mail: tkachuk@ktf.franko.lviv.ua}
\\
  {\small Department of Theoretical Physics, Ivan Franko National University of Lviv,}\\
{\small 12 Drahomanov St., Lviv, UA-79005, Ukraine}}
\begin{document}
\maketitle

\abstract{We investigated the elastic scattering problem with
deformed Heisenberg algebra leading to the existence of a minimal
length. The continuity equations for the moving particle in
deformed space were constructed. We obtained the Green's function
for a free particle, scattering amplitude and cross-section in
deformed space. We also calculated the scattering amplitudes and
differential cross-sections for the Yukawa and the Coulomb
potentials in the Born approximation.  }

\section{Introduction}

 In recent years a number of works were devoted to the investigation of
the quantum mechanical problems with deformed commutation
relations. Such an interest was prompted by several independent
lines of investigation in string theory and quantum gravity which
proposed the existence of a finite lower bound to a possible
resolution of length \cite{gross, maggiore, witten}. Kempf
\textit{et al.} argued that the minimal length can be obtained
from the deformed (generalized) commutation relations
\cite{kempf1,kempf2,kempf3,kempf4,kempf5}. But it should to note
that the deformed algebra leading to a quantized space-time was
originally introduced by Snyder in the relativistic case
\cite{snyder}. The deformed algebra leading to the existence of a
minimal length in $D$-dimensional case reads
\begin{eqnarray}\label{algebra}
\begin{array}{l}
[X_i, P_j]=i\hbar(\delta_{ij}(1+\beta P^2)+\beta'P_iP_j), [P_i,
P_j]=0,
\\
\\
\displaystyle {[X_i,
X_j]}=i\hbar\frac{(2\beta-\beta')+(2\beta+\beta')\beta
P^2}{1+\beta P^2}(P_iX_j-P_jX_i),
\end{array}
\end{eqnarray}
where $\beta$, $\beta'$ are the parameters of deformation. We
assume that these quantities are nonnegative
$\beta,\beta'\geqslant0$. Using the uncertainty relation one can
obtain that minimal length equals $\hbar\sqrt{\beta+\beta'}$.

Deformed Heisenberg algebra (\ref{algebra}) causes new
complications in solving quantum mechanical problems. There just a
few known problems for which the energy spectra have been found
exactly. They are one-dimensional harmonic oscillator with minimal
uncertainty in position \cite{kempf2} and also with minimal
uncertainty in position and momentum \cite{Tkachuk1,Tkachuk2},
$D$-dimensional isotropic harmonic oscillator \cite{chang, Dadic},
three-dimensional relativistic Dirac oscillator \cite{quesne} and
one-dimensional Coulomb problem \cite{fityo}.

The hydrogen atom problem has the crucial role for the
understanding of the key points of modern physics. So it is
interestingly to study this problem when the position and the
momenta operators satisfy the deformed commutation relations
(\ref{algebra}). The Coulomb problem in deformed space with
minimal length was firstly considered by Brau in the special case
$2\beta=\beta'$ \cite{Brau}. The general case of deformation
$2\beta\neq\beta'$ was examined in Ref. \cite{Benczik}. The
authors utilized perturbation theory for calculating of the
corrections to the energy levels. But the perturbation theory used
by the authors did not allow to obtain the corrections for the
$s$-levels. To avoid this problem the authors used the cut-off
procedure and the numerical methods. In our work \cite{mykola} a
modified perturbation theory allowing to calculate corrections for
arbitrary energy levels including $s$-levels was developed. In
\cite{stetsko2} a modified perturbation theory was used for
finding of the corrections to the $ns$-levels of the hydrogen
atom.

It is evidentially that the scattering problem is a key one
because it can also manifest noncommutative effects on the
experimental level. At the same time as far as we know there were
no papers on the investigation of the quantum-mechanical
scattering problem in deformed space with minimal length described
by algebra (\ref{algebra}). There are only a few papers where
scattering problem on the noncommutative space with canonical
deformation $[X_i,X_j]=i\theta_{ij}$ was considered
\cite{Demetrian, Bellucci05,Alavi}.

 In this work a scattering problem in the deformed space with minimal length is studied. We
consider the elastic scattering on Yukawa and Coulomb potentials.
This paper is organized as follows. In the second section we
construct the continuity equation for the moving particle
 in the deformed space. In the third section we obtain the
expressions for the scattering amplitude and the differential
cross-section. In the fourth section we calculate the scattering
amplitude and the cross-section in the Born approximation. And
finally the fifth section contains the discussion.

\section{Continuity equation}
Before considering the scattering problem in deformed space it is
necessary to establish continuity equation at first. The
construction of the continuity equation in deformed space for the
particle moving in the arbitrary external field is more
complicated than in the case of ordinary quantum mechanics. Since
the main goal of our paper is the scattering problem for a
particle in the external Yukawa and Coulomb fields we investigate
the continuity equation for these particular problems only.

Let us consider the particle in deformed space described by the
Hamiltonian
\begin{equation}\label{hamiltonian}
H=\frac{\textbf{P}^2}{2m}+U(R),
\end{equation}
where $U(R)=-e^2\frac{e^{-\lambda R}}{R}$ is the Yukawa potential.
The operators of position $X_i$ and momentum $P_i$ obey the
deformed commutation relations (\ref{algebra}) and
$R=\sqrt{\sum^3_{i=1}X^2_i}$ is the distance operator. The Coulomb
potential can be obtained from the Yukawa potential in the limit
$\lambda=0$.

For constructing the continuity equation we write the
Schr\"{o}dinger equation
\begin{equation}\label{Schroedinger}
i\hbar\frac{\partial\psi}{\partial t}=H\psi.
\end{equation}

One can write the following relation using equation
(\ref{Schroedinger})
\begin{equation}\label{cont1}
\frac{\partial\rho}{\partial t}=\frac{1}{i\hbar}(\psi^*H\psi-\psi
H\psi^*),
\end{equation}
where $\rho=|\psi|^2$ and the Hamiltonian is real: $H^*=H$.

Then for constructing the continuity equation it is necessary to
use the representation of the operators of positions and momenta
satisfying the deformed commutation relations (\ref{algebra}). The
momentum representation for such an algebra is well known, but it
is not convenient for us. We use the following representation that
fulfils algebra (\ref{algebra}) in the first order in $\beta$,
$\beta'$
\begin{eqnarray}\label{rep1}
\left\{
\begin{array}{l}
\displaystyle
X_i=x_i+\frac{2\beta-\beta'}{4}\left(x_ip^2+p^2x_i\right),
\\
\\
\displaystyle P_i=p_i+\frac{\beta'}{2}p_ip^2,
\end{array}
\right.
\end{eqnarray}
where $p^2=\sum^3_{j=1}p^2_j$ and the operators $x_i$, $p_j$
satisfy the canonical commutation relation. The position
representation $x_i=x_i$, $p_j=i\hbar\frac{\partial}{\partial
x_j}$ can be taken for the ordinary Heisenberg algebra.  We notice
that in the special case $2\beta=\beta'$ the position operators
commute in linear approximation over the deformation parameters,
i.e. $[X_i,X_j]=0$.

We rewrite Hamiltonian (\ref{hamiltonian}) using representation
(\ref{rep1}) and develop it with respect to $\beta$, $\beta'$ up
to the first order. The main problem is related with the expansion
of the distance operator $R$ and inverse distance operator $1/R$.
In our previous paper \cite{mykola} for the elimination of
divergent term $1/r$ at $r=0$ in the expansion of $R$ and as a
consequence $1/r^3$ in $1/R$ we proposed the so-called shifted
expansion. So for $R$ we have \cite{mykola}
\begin{equation}\label{distance_operator}
R=\sqrt{r^2+b^2}+\frac{\alpha}{2}(rp^2+p^2r),
\end{equation}
where $\alpha=(2\beta-\beta')/2$ and $b=\hbar\sqrt{2\alpha}$.

Using expansion (\ref{distance_operator}) we can represent
decomposition for the inverse distance operator just as it was
done in \cite{mykola}
\begin{equation}\label{inverse_distance}
\frac{1}{R}=\frac{1}{\sqrt{r^2+b^2}}-\frac{\alpha}{2}\left(\frac{1}{r}p^2+p^2\frac{1}{r}\right).
\end{equation}

For the expansion of the exponential operator we use the
$T$-exponent representation
\begin{equation}\label{T_exponent}
e^{-\lambda\left[\sqrt{r^2+b^2}+\frac{\alpha}{2}(rp^2+p^2r)\right]}=
e^{-\lambda\sqrt{r^2+b^2}}T\exp{\left(\frac{\alpha}{2}\int^{\lambda}_0\rm{d}\lambda'e^{\lambda'\sqrt{r^2+b^2}}
(rp^2+p^2r)e^{-\lambda'\sqrt{r^2+b^2}}\right)}.
\end{equation}

In this paper we restrict ourselves to the first order
approximation over the deformation parameters. So we develop the
$T$-exponent operator into the series and take into account only
the first order terms in $\alpha$.  Then we use the explicit form
for the operator $p^2=-\hbar^2\nabla^2$ and after simple
transformations we represent expression (\ref{T_exponent}) in the
following way
\begin{eqnarray}\label{exponent_decomposition_integrated}
\begin{array}{c}
\displaystyle
e^{-\lambda\left[\sqrt{r^2+b^2}+\frac{\alpha}{2}(rp^2+p^2r)\right]}=e^{-\lambda\sqrt{r^2+b^2}}\left(1+
\frac{\alpha\hbar^2}{2}\int^{\lambda}_0\rm{d}\lambda'\left[2\lambda'^2r-2\lambda'
((\textbf{r}\nabla)+(\nabla\textbf{r}))+r\nabla^2+\nabla^2r\right]\right)\simeq
\\
\\
\displaystyle e^{-\lambda\sqrt{r^2+b^2}}+\frac{\alpha\hbar^2}{2}
e^{-\lambda
r}\left(\frac{2}{3}\lambda^3r-\lambda^2((\textbf{r}\nabla)+(\nabla\textbf{r}))+\lambda(r\nabla^2+\nabla^2r)\right).
\end{array}
\end{eqnarray}

Using expansion (\ref{inverse_distance}) for the inverse distance
and decomposition (\ref{exponent_decomposition_integrated}) we
represent the Yukawa potential in the form of
\begin{eqnarray}\label{Yukawa_decomposed}
\begin{array}{c}
\displaystyle U(R)=-e^2\frac{e^{-\lambda R}}
{R}=-e^2\left[\frac{e^{-\lambda\sqrt{r^2+b^2}}}{\sqrt{r^2+b^2}}+\frac{\alpha\hbar^2}{2}e^{-\lambda
r}\left(\frac{1}{r}\nabla^2+\nabla^2\frac{1}{r}\right)+\frac{\alpha\hbar^2}{2}\frac{e^{-\lambda
r}}{r}\left(\frac{2}{3}\lambda^3r-\right.\right.
\\
\\
\displaystyle
\left.\left.\lambda^2((\textbf{r}\nabla)+(\nabla\textbf{r}))+\lambda(r\nabla^2+\nabla^2r)\right)
+\frac{\alpha\hbar^2}{2}e^{-\lambda
r}\left(\frac{2\lambda^2}{r}-2\lambda\frac{1}{r}((\textbf{r}\nabla)+(\nabla\textbf{r}))\frac{1}{r}\right)\right]=
U(\textbf{r},\textbf{p})
\end{array}
\end{eqnarray}
and we note that here $\textbf{p}=-i\hbar\nabla$.

Thus Hamiltonian (\ref{hamiltonian}) in canonical variables reads
\begin{equation}\label{hamiltonian_decomposed}
H=\frac{p^2}{2m}+\frac{\beta'p^4}{2m}+U(\textbf{r},\textbf{p}).
\end{equation}

We substitute Hamiltonian (\ref{hamiltonian_decomposed}) in
relation (\ref{cont1}) and take into consideration the explicit
form for the operator $p^2=-\hbar^2\nabla^2$. After simple
transformations we obtain
\begin{eqnarray}\label{continuity_Yukawa}
\begin{array}{c}
\displaystyle\frac{\partial\rho}{\partial
t}=\frac{1}{i\hbar}\nabla\left[-\frac{\hbar^2}{2m}(\psi^*\nabla\psi-\psi\nabla\psi^*)+\frac{\beta'\hbar^4}{2m}
(\psi^*\nabla^3\psi-\psi\nabla^3\psi^*-\nabla\psi^*\nabla^2\psi+\nabla\psi\nabla^2\psi^*)
-\frac{e^2\hbar^2(2\beta-\beta')}{4}\times\right.
\\
\\
\displaystyle\left.\left(\psi^*\left[e^{-\lambda
r}\left(\frac{1}{r}\nabla+\nabla\frac{1}{r}\right)+\lambda\frac{e^{-\lambda
r}}{r}(r\nabla+\nabla r)\right]\psi-\psi\left[e^{-\lambda
r}\left(\frac{1}{r}\nabla+\nabla\frac{1}{r}\right)+\lambda\frac{e^{-\lambda
r}}{r}(r\nabla+\nabla r)\right]\psi^*\right)\right],
\end{array}
\end{eqnarray}
where $\nabla^3=(\nabla,\nabla)\nabla$.

Relation (\ref{continuity_Yukawa}) can be represented in the
following form
\begin{equation}
\frac{\partial\rho}{\partial t}+\rm{div}\textbf{j}=0.
\end{equation}
So we have a well known continuity equation where
\begin{eqnarray}\label{current_Yukawa}
\begin{array}{c}
\displaystyle\textbf{j}=-\frac{1}{i\hbar}\left[-\frac{\hbar^2}{2m}(\psi^*\nabla\psi-\psi\nabla\psi^*)+\frac{\beta'\hbar^4}{2m}
(\psi^*\nabla^3\psi-\psi\nabla^3\psi^*-\nabla\psi^*\nabla^2\psi+\nabla\psi\nabla^2\psi^*)
-\frac{e^2\hbar^2(2\beta-\beta')}{4}\times\right.
\\
\\
\displaystyle\left.\left(\psi^*\left[e^{-\lambda
r}\left(\frac{1}{r}\nabla+\nabla\frac{1}{r}\right)+\lambda\frac{e^{-\lambda
r}}{r}(r\nabla+\nabla r)\right]\psi-\psi\left[e^{-\lambda
r}\left(\frac{1}{r}\nabla+\nabla\frac{1}{r}\right)+\lambda\frac{e^{-\lambda
r}}{r}(r\nabla+\nabla r)\right]\psi^*\right)\right]
\end{array}
\end{eqnarray}
is the density current for the particle in the external Yukawa
field.

In a particular case of $\lambda=0$ we obtain the density current
for a particle in the Coulomb field
\begin{eqnarray}\label{current_Coulomb1}
\begin{array}{c}
\displaystyle\textbf{j}=\frac{1}{i\hbar}\left(-\frac{\hbar^2}{2m}(\psi^*\nabla\psi-\psi\nabla\psi^*)+
\frac{\beta'\hbar^4}{2m}(\psi^*\nabla^3\psi-\psi\nabla^3\psi^*-\nabla\psi^*\nabla^2\psi+\nabla\psi\nabla^2\psi^*)-\right.
\\
\\
\displaystyle\left.\frac{\hbar^2e^2(2\beta-\beta')}{4}\left(\psi^*\left(\frac{1}{r}\nabla+\nabla\frac{1}{r}\right)\psi
-\psi\left(\frac{1}{r}\nabla+\nabla\frac{1}{r}\right)\psi^*\right)\right).
\end{array}
\end{eqnarray}

Expressions (\ref{current_Yukawa}) and (\ref{current_Coulomb1})
for the density current in the deformed case for a particle moving
in the external Yukawa or Coulomb fields are somewhat different
from the density current in the ordinary quantum mechanics: we
have two additional terms into the continuity equation. One of
them is caused by the deformed kinetic energy. The second
contribution is caused by the external fields. We notice that in
the special case of $2\beta=\beta'$ when the position operators
commute in linear approximation over the deformation parameters,
i.e. $[X_i,X_j]=0$, the potential energy does not give any
contribution into the continuity equation.

But we want to stress that in the scattering problems we calculate
the density current at large distances from the scattering center
and we can neglect the terms caused by the field. So the density
current for a particle scattered by the Yukawa or the Coulomb
fields at large distances from the scattering center takes the
same form as for a free particle in the deformed space
\begin{equation}\label{density_current_scattering}
\textbf{j}=\frac{1}{i\hbar}\left(-\frac{\hbar^2}{2m}(\psi^*\nabla\psi-\psi\nabla\psi^*)+
\frac{\beta'\hbar^4}{2m}(\psi^*\nabla^3\psi-\psi\nabla^3\psi^*-\nabla\psi^*\nabla^2\psi+\nabla\psi\nabla^2\psi^*)\right).
\end{equation}

\section{Scattering amplitude}

In this section we examine the scattering of the particle on the
arbitrary potential $U(\textbf{R})$. Position operators in the
potential energy operator fulfil the deformed Heisenberg algebra
(\ref{algebra}). Using representation (\ref{rep1}) the potential
energy operator can be represented as a function of the canonical
variables $x_i,p_j$: $U(\textbf{R})=U(\textbf{r},\textbf{p})$ (see
eq. (\ref{Yukawa_decomposed})). The explicit form of the potential
is taken into account in the final relations only.

Let us consider the Schr\"{o}dinger equation
\begin{equation}\label{Schroedinger2}
\left(\frac{p^2}{2m}+\frac{\beta'p^4}{2m}+U(\textbf{r},\textbf{p})\right)\Psi=E\Psi.
\end{equation}
We suppose that $U(\textbf{r},\textbf{p})\rightarrow 0$ when
$r\rightarrow \infty$ and at large distances from the scatterer
the motion of a particle is free.
 The kinetic energy of the incident particle equals
\begin{equation}\label{kinetic_energy}
E=\frac{\hbar^2k^2}{2m}(1+\beta'\hbar^2k^2),
\end{equation}
 where $\textbf{k}$ is
the wave vector of an incident particle and
$\textbf{P}=\hbar\textbf{k}(1+\beta'\hbar^2k^2)$ is the momentum
of a particle. The wave function of incident particle is
\begin{equation}\label{plane_wave}
\psi_k(\textbf{r})=e^{i{\bf kr}}.
\end{equation}

After scattering the motion of a particle is also free at long
distances from the scatterer with the momentum
$\textbf{P}'=\hbar\textbf{k}'(1+\beta'\hbar^2k'^2)$. As was noted
above we consider elastic scattering so we have $k'=k$.

Equation (\ref{Schroedinger2}) can be represented as follows
\begin{equation}\label{eq1}
(\nabla^2-\beta'\hbar^2\nabla^4+k^2[1+\beta'\hbar^2k^2])\Psi=\frac{2m}{\hbar^2}U(\textbf{r},\textbf{p})\Psi
\end{equation}
and as was noticed earlier $\textbf{p}=-i\hbar{\bf \nabla}$.

The formal solution of equation (\ref{eq1}) can be written as
\begin{equation}\label{solution}
\Psi(\textbf{r})=\psi_k(\textbf{r})+\int
G(\textbf{r},\textbf{r}')\frac{2m}{\hbar^2}U(\textbf{r}',\textbf{p}')\Psi(\textbf{r}')\textrm{d}\textbf{r}',
\end{equation}
that in fact is the integral equation and
$G(\textbf{r},\textbf{r}')$ is the Green's function which
satisfies the following equation
\begin{equation}\label{Green}
(\nabla^2-\beta'\hbar^2\nabla^4+k^2[1+\beta'\hbar^2k^2])G(\textbf{r},\textbf{r}')=\delta(\textbf{r}-\textbf{r}').
\end{equation}

The solution for the Green's function reads
\begin{equation}
G(\textbf{r},\textbf{r}')=G(\textbf{r}-\textbf{r}')=\frac{1}{(2\pi)^3}\int
\frac{e^{i{\bf q(r-r')
}}}{k^2(1+\beta'\hbar^2k^2)-q^2(1+\beta'\hbar^2q^2)}\textrm{d}\textbf{q}.
\end{equation}
After integration by angular variables we have
\begin{equation}\label{int1}
G(\textbf{r},\textbf{r}')=\frac{1}{4i\pi^2|\textbf{r}-\textbf{r}'|}\int\limits^{+\infty}_{-\infty}
\frac{qe^{iq|{\bf r-r'}|}}
{k^2(1+\beta'\hbar^2k^2)-q^2(1+\beta'\hbar^2q^2)}\textrm{d}q.
\end{equation}
The last integral can be calculated using the calculus of
residues. Therefore, it is necessary to determine the contour of
integration in the complex plane. So the way of enclosing the
poles $q=\pm k$ might be defined. The manner of enclosing the
poles can be determined from the boundary conditions imposed on
the Green's function $G(|\textbf{r}-\textbf{r}'|)$ when
$|\textbf{r}-\textbf{r}'|\rightarrow 0$. For obtaining the
solution corresponding to the outgoing wave we choose the contour
of integration similarly as we select the contour for the outgoing
wave in ordinary quantum mechanics. Then it is easy to calculate
integral (\ref{int1}) and we have
\begin{equation}
G(|\textbf{r}-\textbf{r}'|)=-\frac{1}{4\pi|{\bf r-r'
}|(1+2\beta'\hbar^2k^2)} \left(e^{ik|{\bf r-r'
}|}-e^{-\sqrt{k^2+1/(\beta'\hbar^2)}|{\bf r-r'}|}\right).
\end{equation}
Green's function for a free particle in the deformed case is
somewhat different than in the ordinary one. We have an additional
multiplier $1/(1+2\beta'\hbar^2k^2)$ tending to unity when the
$\beta'\rightarrow 0$. Then in the deformed case we also have the
additional function $e^{-\sqrt{k^2+1/(\beta'\hbar^2)}|{\bf
r-r'}|}$ rapidly decreasing with increasing
$|\textbf{r}-\textbf{r}'|$. The last function tends to zero when
$\beta'\rightarrow 0$. Since we want to find the asymptotic
behaviour for the Green's function taking place when the
difference $|\textbf{r}-\textbf{r}'|$ is large  we may neglect the
last function. Thus the Green's function at large distances from
the scattering center reads
\begin{equation}\label{GreenSimplified}
G(|\textbf{r}-\textbf{r}'|)=-\frac{1}{4\pi|\textbf{r}-\textbf{r}'|(1+2\beta'\hbar^2k^2)}
e^{ik|{\bf r-r'}|}.
\end{equation}

Using this Green's function we can rewrite integral equation
(\ref{solution}) in the form
\begin{equation}\label{inteq}
\Psi(\textbf{r})=\psi_k(\textbf{r})-\frac{m}{2\pi\hbar^2(1+2\beta'\hbar^2k^2)}\int
\frac{e^{ik|{\bf r-r'}|}}{|\textbf{r}-\textbf{r}'|}
U(\textbf{r}',\textbf{p}')\Psi(\textbf{r}')\textrm{d}\textbf{r}'.
\end{equation}

As was noted earlier we want to obtain the solution of integral
equation (\ref{inteq}) at large distances from the center of
scatter. The action of the potential energy operator
$U(\textbf{r}',\textbf{p}')$ on the wave function makes a
considerable contribution to integral of equation (\ref{inteq}) in
the bounded domain. The action of the operator
$U(\textbf{r}',\textbf{p}')$ on the wave function
$\Psi(\textbf{r}')$ makes a negligibly small contribution to
integral (\ref{inteq}) for larger $r'$. So the actual values of
the integration variable $\textbf{r}'$ are bounded by some
effective radius of the potential energy. When
$r\rightarrow\infty$ the ratio $r'/r$ is small and one can use the
following development
\begin{equation}\label{decomposition}
|\textbf{r}-\textbf{r}'|=\sqrt{r^2-2\textbf{r}\textbf{r}'+r'^2}\simeq
r\left(1-\frac{\textbf{r}\textbf{r}'}{r^2}+\cdots\right)=r-\textbf{n}\textbf{r}'+\cdots,
\end{equation}
where $\textbf{n}=\textbf{r}/r$ is the unit vector along the
direction of motion of the scattered particle.

Then we substitute decomposition (\ref{decomposition}) in equation
(\ref{inteq}) and taking into account the leading term of the
asymptotic we represent equation (\ref{inteq}) in the form
\begin{equation}\label{wavefunction}
\Psi(\textbf{r})=\psi_k(\textbf{r})-\frac{m}{2\pi\hbar^2(1+2\beta'\hbar^2k^2)}\frac{e^{ikr}}{r}\int
e^{-i\textbf{k}'\textbf{r}'}U(\textbf{r}',\textbf{p}')\Psi(\textbf{r}')\textrm{d}\textbf{r}',
\end{equation}
where $\textbf{k}'=k\textbf{n}$.

Expression (\ref{wavefunction}) can be rewritten as follows
\begin{equation}\label{WaveFunct2}
\Psi(\textbf{r})=e^{i{\bf kr}}+\frac{e^{ikr}}{r}f,
\end{equation}
where
\begin{equation}\label{scattering_amplitude}
f=-\frac{m}{2\pi\hbar^2(1+2\beta'\hbar^2k^2)}\int e^{-i{\bf k'r'
}}U(\textbf{r}',\textbf{p}')\Psi(\textbf{r}')\textrm{d}\textbf{r}'
\end{equation}
is the scattering amplitude.

Expression (\ref{WaveFunct2}) formally coincides with the wave
function of the scattering problem in ordinary quantum mechanics.
The main difference between them is caused by different
expressions for the scattering amplitude. As in the ordinary case
the second term of the wave function (\ref{WaveFunct2})
corresponds to the wave function of the scattered particle
\begin{equation}\label{outgoing_wave}
\psi_{\rm scatt}=\frac{e^{ikr}}{r}f.
\end{equation}

The central problem of the scattering theory is the calculation of
the differential cross-section. We define the differential
cross-section similarly as it was done in the ordinary quantum
mechanics
\begin{equation}
d\sigma=\frac{\textbf{j}_{\rm scatt}d\textbf{S}}{j_0},
\end{equation}
where $j_0$ is the absolute value of the current density for the
incident particles, $\textbf{j}_{scatt}$ is the current density
for the scattered particles and $d\textbf{S}$ is the element of
the area along the direction of motion for the scattered
particles. The element of  the area can be rewritten in the form
$d\textbf{S}=\textbf{n}dS$ where $\textbf{n}=\textbf{r}/r$. After
introducing a spatial angle $d\Omega=dS/r^2$ the differential
cross-section can be represented in the form
\begin{equation}\label{diff_cross-section}
\frac{d\sigma}{d\Omega}=\frac{(\textbf{j}_{\rm
scatt}\textbf{n})r^2}{j_0}.
\end{equation}

For the calculation of the differential cross-section it is
necessary to calculate the current density for the incident
particles and the current density for the scattered particles.
Substituting (\ref{plane_wave}) into
(\ref{density_current_scattering}) we obtain
\begin{equation}\label{current_free_particle}
\textbf{j}_0=\frac{\hbar\textbf{k}}{m}(1+2\beta'\hbar^2k^2).
\end{equation}
The expression for the current density of incident particles is
somewhat different than in the ordinary case. We have the
additional factor $(1+2\beta'\hbar^2k^2)$ tending to unity if
$\beta'\rightarrow 0$.

Substituting function (\ref{outgoing_wave}) in relation
(\ref{density_current_scattering}) and after simple
transformations we obtain the following expression for the current
density for the outgoing particles
\begin{equation}\label{current_Coulomb_simplified}
\textbf{j}_{\rm scatt}=\frac{\hbar\textbf{k}}{m}\frac{|f|^2}{r^2}
(1+2\beta'\hbar^2k^2).
\end{equation}
Then we substitute relations (\ref{current_Coulomb_simplified})
and (\ref{current_free_particle}) in expression
(\ref{diff_cross-section}) and obtain
\begin{equation}\label{cross-section_Coulomb}
\frac{d\sigma}{d\Omega}=|f|^2.
\end{equation}

As we see the relation for the differential cross-section formally
coincides with the expression for the cross-section in ordinary
quantum mechanics, but the scattering amplitude is defined by
relation (\ref{scattering_amplitude}) which, as we noted before,
is somewhat different from the ordinary one.

\section{Born approximation}
The Born approximation can be used for finding the scattering
amplitude in the deformed case. In the Born approximation we
consider the potential energy $U(\textbf{r},\textbf{p})$ as a
small perturbation and the integral equation (\ref{wavefunction})
can be solved by the method of successive approximation. In the
first approximation we substitute the plane wave
(\ref{plane_wave}) in expression (\ref{scattering_amplitude}) and
taking into account the explicit form for the Yukawa potential
(\ref{Yukawa_decomposed}) we obtain scattering amplitude
\begin{eqnarray}\label{Yukawa_amplitude}
\begin{array}{c}
\displaystyle f_{\rm
Yukawa}=\frac{m}{2\pi\hbar^2(1+2\beta'\hbar^2k^2)}\displaystyle{\int}
e^{-i{\bf k'r
}}e^2\left[\frac{e^{-\lambda\sqrt{r^2+b^2}}}{\sqrt{r^2+b^2}}+
\frac{\hbar^2(2\beta-\beta')}{4}e^{-\lambda
r}\left(\frac{1}{r}\nabla^2+\nabla^2\frac{1}{r}\right)+\frac{\hbar^2(2\beta-\beta')}{4}\frac{e^{-\lambda
r}}{r}\times\right.
\\
\\
\displaystyle\left.\left(\frac{2}{3}\lambda^3r-\lambda^2((\textbf{r}\nabla)+(\nabla\textbf{r}))+\lambda(r\nabla^2+\nabla^2r)\right)
+\frac{\hbar^2(2\beta-\beta')}{4}e^{-\lambda
r}\left(\frac{2\lambda^2}{r}-2\lambda\frac{1}{r}((\textbf{r}\nabla)+(\nabla\textbf{r}))\frac{1}{r}\right)\right]
e^{i{\bf kr}}\textrm{d}{\bf r}.
\end{array}
\end{eqnarray}

Then we calculate the contribution into the scattering amplitude
caused by the term
$e^{-\lambda\sqrt{r^2+b^2}}\over{\sqrt{r^2+b^2}}$
\begin{equation}\label{Integral_root}
\int e^{-i{\bf
k'r}}e^2\frac{e^{-\lambda\sqrt{r^2+b^2}}}{\sqrt{r^2+b^2}} e^{i{\bf
kr}}d\textbf{r}=\frac{4\pi
e^2b}{\sqrt{\lambda^2+q^2}}K_1(b\sqrt{\lambda^2+q^2}),
\end{equation}
where $K_1$ is the Bessel function \cite{abramowitz} and
\begin{equation}\label{wave_number}
q=|\textbf{k}'-\textbf{k}|=2k\sin{\frac{\vartheta}{2}},
\end{equation}
$\vartheta$ is the scattering amplitude.

We develop the Bessel function into the series and take into
account only the first order terms in $b$. So contribution
(\ref{Integral_root}) can be represented in the form
\begin{eqnarray}
\begin{array}{c}
\displaystyle\frac{4\pi
e^2b}{\sqrt{\lambda^2+q^2}}K_1(b\sqrt{\lambda^2+q^2})\simeq\frac{4\pi
e^2b}{\sqrt{\lambda^2+q^2}}\left(\frac{1}{b\sqrt{\lambda^2+q^2}}+\frac{b\sqrt{\lambda^2+q^2}}{2}
\left(\ln\left(\frac{b\sqrt{\lambda^2+q^2}}{2}\right)+\gamma-\frac{1}{2}\right)\right)=
\\
\\
\displaystyle=\frac{4\pi e^2}{\lambda^2+q^2}+2\pi
e^2b^2\left(\ln\left(\frac{b\sqrt{\lambda^2+q^2}}{2}\right)+\gamma-\frac{1}{2}\right),
\end{array}
\end{eqnarray}
where $\gamma=0.57721...$ is the Euler constant.

It is easy to calculate the contributions caused by another terms
in integral (\ref{Yukawa_amplitude}). So we can write the
scattering amplitude for the Yukawa potential taking into account
the explicit form for the parameter $b=\hbar\sqrt{2\beta-\beta'}$
\begin{eqnarray}\label{amplitude_Yukawa_integrated}
\begin{array}{c}
\displaystyle f_{\rm
Yukawa}=\frac{me^2}{2\pi\hbar^2(1+2\beta'\hbar^2k^2)}\left(\frac{4\pi
}{\lambda^2+q^2}+\pi\hbar^2
(2\beta-\beta')\left(\ln\left(\frac{\hbar^2(2\beta-\beta')(\lambda^2+q^2)}{4}\right)+2\gamma-1\right)-\right.
\\
\\
\displaystyle\left.2\pi\hbar^2(2\beta-\beta')\left[\frac{k^2}{\lambda^2+q^2}+
\frac{\lambda^2}{(\lambda^2+q^2)^2}\left(2k^2-\frac{\lambda^2}{3}\right)\right]\right).
\end{array}
\end{eqnarray}

Since we take into consideration only the first order terms in
$\beta$, $\beta'$ we can rewrite the last expression in the form
\begin{eqnarray}\label{amplitude_Yukawa_linearized}
\begin{array}{c}
\displaystyle f_{\rm
Yukawa}=\frac{2me^2}{\hbar^2(\lambda^2+q^2)}+\frac{me^2}{2}(2\beta-\beta')\left[
\ln{\left(\frac{\hbar^2(2\beta-\beta')(\lambda^2+q^2)}{4}\right)}+\right.
\\
\\
\displaystyle\left.2\gamma-1-
\frac{2k^2}{\lambda^2+q^2}-\frac{2\lambda^2}{(\lambda^2+q^2)^2}\left(2k^2-\frac{\lambda^2}{3}\right)
\right]-\beta'\frac{4me^2k^2}{\lambda^2+q^2}.
\end{array}
\end{eqnarray}

Using the scattering amplitudes for the Yukawa potential
(\ref{amplitude_Yukawa_linearized}) and taking into account
relation (\ref{wave_number}) we calculate the differential
cross-sections
\begin{eqnarray}\label{diff_cross-sect_Yukawa}
\begin{array}{c}
\displaystyle\frac{d\sigma}{d\Omega}=\frac{4m^2e^4}{\hbar^4\left(\lambda^2+4k^2\sin^2\frac{\vartheta}{2}\right)^2}
+\frac{4me^2}{\hbar^2\left(\lambda^2+4k^2\sin^2{\frac{\vartheta}{2}}\right)}\left(\frac{me^2}{2}(2\beta-\beta')\left[
\ln\left(\frac{\hbar^2(2\beta-\beta')\left(\lambda^2+4k^2\sin^2\frac{\vartheta}{2}\right)}{4}\right)+\right.
\right.
\\
\\
\displaystyle\left.\left.2\gamma-1-\frac{2k^2}{\lambda^2+4k^2\sin^2\frac{\vartheta}{2}}-
\frac{2\lambda^2}{\left(\lambda^2+4k^2\sin^2\frac{\vartheta}{2}\right)^2}\left(2k^2-\frac{\lambda^2}{3}\right)\right]-\beta'\frac{4me^2k^2}
{\lambda^2+4k^2\sin^2\frac{\vartheta}{2}}\right).
\end{array}
\end{eqnarray}
Putting $\lambda=0$ we obtain the differential cross-section for
the Coulomb potential
\begin{eqnarray}\label{diff_cross-sect_Coulomb}
\begin{array}{c}
\displaystyle\frac{d\sigma}{d\Omega}=\frac{m^2e^4}{4\hbar^4k^4\sin^4\frac{\vartheta}{2}}+
\frac{me^2}{\hbar^2k^2\sin^2\frac{\vartheta}{2}}\left(\frac{me^2}{2}(2\beta-\beta')
\left[\ln\left(\hbar^2(2\beta-\beta')k^2\sin^2\frac{\vartheta}{2}\right)+\right.\right.
\\
\\
\displaystyle\left.\left.2\gamma-1-
\frac{1}{2\sin^2\frac{\vartheta}{2}}\right]-\beta'\frac{me^2}{\sin^2\frac{\vartheta}{2}}\right).
\end{array}
\end{eqnarray}

Similarly to \cite{Benczik,mykola,stetsko2} we introduce two
parameters $\Delta x_{\min}=\hbar\sqrt{\beta+\beta'}$ (or
$\xi=\Delta x_{\min}/a$) and $\eta=\frac{\beta}{\beta+\beta'}$
instead of $\beta$ and $\beta'$. As was noted in \cite{mykola}
that our calculations hold if $2\beta-\beta'\geqslant 0$ and
$\beta$, $\beta'$ are nonnegative constants. These conditions lead
to the constraints on the domain of variation for the
dimensionless parameter $\eta$: $\frac{1}{3}\leqslant\eta\leqslant
1$. The last expression for the differential cross-section on the
Coulomb potential can be rewritten using the parameters $\Delta
x_{\min}$ and $\eta$ as follows
\begin{equation}\label{cross-section_rewritten}
\frac{d\sigma}{d\Omega}=\frac{m^2e^4}{4\hbar^4k^4\sin^4\frac{\vartheta}{2}}\left[1+\zeta\left(\Delta
x_{\min}, \eta, k, \vartheta \right)\right],
\end{equation}
where
\begin{eqnarray}\label{zeta_funct}
\begin{array}{c}
\zeta\left(\Delta x_{\min}, \eta, k, \vartheta\right)=2\Delta
x^2_{\min}k^2(3\eta-1)\sin^2{\frac{\vartheta}{2}}\left[\ln\left(\Delta
x^2_{\min}k^2(3\eta-1)\sin^2{\frac{\vartheta}{2}}\right)\right.
\\
\\
\left.+2\gamma-1\right]+\Delta x^2_{\min}k^2(\eta-3)
\end{array}
\end{eqnarray}
is the specially introduced function which shows a correction to
the differential cross-section caused by the deformation in terms
of the wave vector.

We represent a relation for the differential cross-section
(\ref{cross-section_rewritten}) in terms of energy of the incident
particle. We emphasize that in the deformed case kinetic energy is
given by relation (\ref{kinetic_energy}). Using
(\ref{kinetic_energy}) and taking into account only linear terms
over deformation parameters we obtain
\begin{equation}
\frac{d\sigma}{d\Omega}=\frac{e^2}{16E^2\sin^4{\frac{\vartheta}{2}}}
\left[1+\delta(\Delta x_{\min},\eta,E,\vartheta)\right],
\end{equation}
where
\begin{eqnarray}\label{cross-sect_energy}
\begin{array}{c}
\delta(\Delta x_{\min},\eta,E,\vartheta)=\zeta(\Delta
x_{\min},\eta,E,\vartheta)+\frac{4m}{\hbar^2}\Delta
x^2_{\min}E(1-\eta)=\frac{4m}{\hbar^2}\Delta
x^2_{\min}E(3\eta-1)\sin^2{\frac{\vartheta}{2}}\times
\\
\\
\left[\ln\left(\frac{2m}{\hbar^2}\Delta
x^2_{\min}E(3\eta-1)\sin^2{\frac{\vartheta}{2}}\right)+
2\gamma-1\right]-\frac{2m}{\hbar^2}\Delta x^2_{\min}E(\eta+1)
\end{array}
\end{eqnarray}
is the specially introduced function which similarly to
(\ref{zeta_funct}) shows the corrections to the cross-section but
expresses it in terms of energy.

Having expression (\ref{cross-sect_energy}) we can numerically
estimate the correction to the cross-section caused by the minimal
length effects. Relation (\ref{cross-sect_energy}) shows that
function $\delta$ depends on the four variables $\Delta x_{\min}$,
$\eta$, $E$ and $\vartheta$.

 For the minimal length we use the upper bound obtained in \cite{Benczik,mykola,stetsko2}.
As was shown in these works the upper bound for the minimal length
is of the order $10^{-16}\div 10^{-17}$m. For the calculating of
$\delta$ we take $\Delta x_{\min}=10^{-16}$m. The parameter $\eta$
have an arbitrary value from its domain of variation.

\begin{figure}
  \centerline{\includegraphics[scale=1.4,clip]{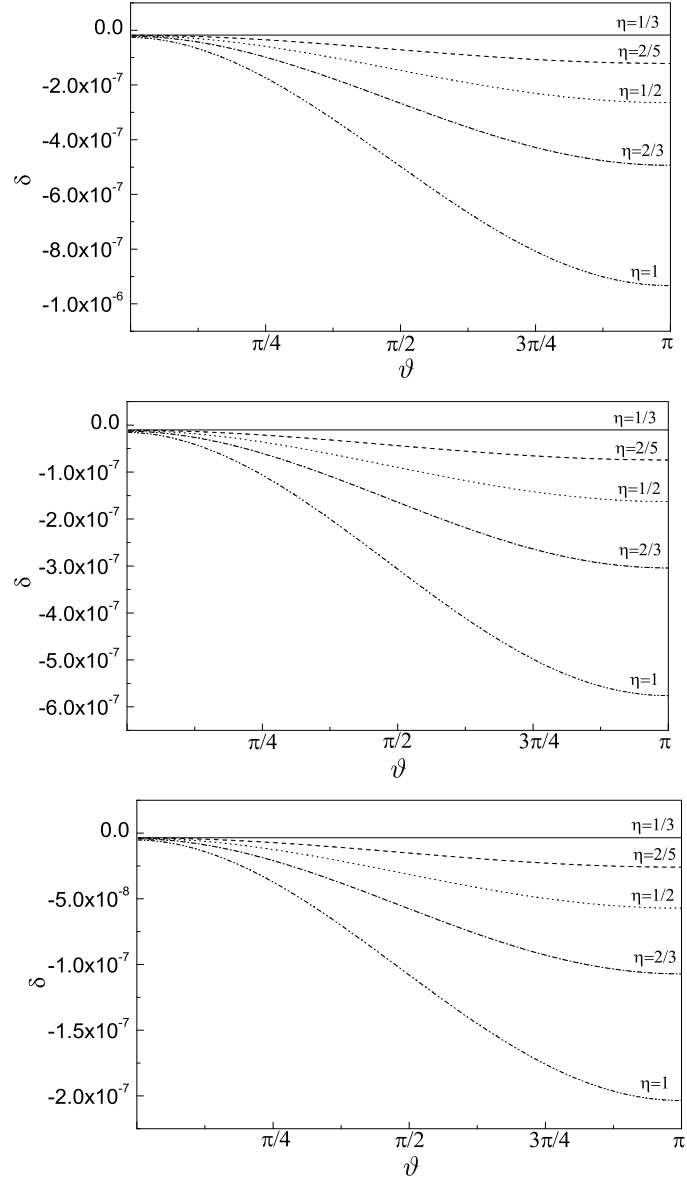}}
  \caption{The dependence of $\delta(\Delta x_{\min},E,\eta,\vartheta)$ on the the scattering angle
  $\vartheta$ for different $\eta$ and energy $E$ of the
  incident particles. The upper graph corresponds to the energy
    of the incident particles $E=50$keV. The middle graph corresponds to the energy $E=30$keV and the lower graph
    corresponds to the energy $E=10$keV.}\label{fig1}
\end{figure}

Fig.\ref{fig1} shows the dependence of $\delta$ on the scattering
angle for different $\eta$ and the energies of incident particles.
We have three graphs and each of them corresponds to the given
energy of incident particles. The function $\delta$ is negative
for these energies and this means that the differential
cross-section in the deformed case is smaller than in ordinary
quantum mechanics. We emphasize that in the special case
$\eta=\frac{1}{3}$ the function $\delta$ does not depend on the
scattering angle and is fully determined by the energy of
particle. When $\eta\neq\frac{1}{3}$ the function $\delta$
decreases with the increasing of the scattering angle.

% We would also like to stress that for our estimations we
%take a sufficiently high energies of the incident particles
%because the Born approximation for the Coulomb potential is good
%at high energies.

\section{Discussion}

We studied the elastic scattering problem for the Yukawa and
Coulomb potentials in the space with deformed Heisenberg algebra
leading to nonzero minimal length. Using the shifted expansion
over the deformation parameters $\beta$ and $\beta'$ we found the
continuity equation in linear approximation over these parameters.
The explicit expression for the density current contains new terms
(in comparison with the undeformed case) caused by deformation.
The terms proportional to $\beta'$ are caused by kinetic energy in
deformed space and the terms proportional to $2\beta-\beta'$ are
caused by potential energy. It is necessary to stress that in
contrast to the ordinary quantum mechanics in the case of
deformation the potential energy also makes a contribution to the
expression for the density current (see eq.
(\ref{current_Yukawa}),(\ref{current_Coulomb1})). For the free
particle the expression for the density current contains only an
additional term proportional to $\beta'$ (see eq.
(\ref{density_current_scattering})).

We obtained the Green's function for the free particle in the
deformed space. Using this Green's function we obtained the wave
function of the scattering problem in the deformed space.
Similarly to the ordinary quantum mechanics the wave function
consists of two terms. One of them is the plane wave corresponding
to the wave function of the incident free particle and the second
is the divergent spherical wave which corresponds to the scattered
outgoing particle. We also obtained the relation for the
scattering amplitude in the deformed space. As was shown the
expression for the scattering amplitude in the deformed case is
similar to the ordinary one but with an additional multiplier
$1/(1+2\beta'\hbar^2k^2)$ tending to the unity when
$\beta'\rightarrow 0$. Using the relation for the scattering
amplitude we found the expression for the differential
cross-section.

We calculated the scattering amplitude for the Yukawa potential in
the Born approximation. Using this expression we found the
differential cross-section for the Yukawa and Coulomb potentials
in the Born approximation. Then we numerically estimate the
corrections to the cross-section for the Coulomb (as a limit of
the Yukawa one) potential caused by the minimal length effects.
For calculations we used the estimations of minimal length
obtained in the works \cite{Benczik, mykola,stetsko2}. We revealed
that in the deformed case the differential cross-section is
smaller than in the ordinary one. Absolute value of corrections to
the cross-section caused deformation increases with the increasing
of energy and scattering angle.

\end{document}